\documentclass[aps,prb,showpacs,superscriptaddress,twocolumn]{revtex4}
\usepackage[latin1]{inputenc}
\usepackage{graphicx}
\usepackage{color}

\newcommand{\pznpt}{(1-x)PZN-xPT~}
\newcommand{\pmnpt}{(1-x)PMN-xPT~}
\newcommand{\pmn}{PbMg$_{1/3}$Nb$_{2/3}$O$_3$~}
\newcommand{\pzn}{PbZn$_{1/3}$Nb$_{2/3}$O$_3$~}

\begin{document} 


\title{Evolution of the neutron quasi-elastic scattering through the 
ferroelectric phase transition in 93\%\pzn - 7\% PbTiO$_3$}


 
\author{G.-M. Rotaru}
\email[]{gelu.rotaru@psi.ch}
\affiliation{Laboratory for Neutron Scattering ETHZ \& Paul Scherrer
Institut, CH-5232 Villigen PSI, Switzerland} 

\author{S.N. Gvasaliya}
\affiliation{Laboratory for Neutron Scattering ETHZ \& Paul Scherrer
Institut, CH-5232 Villigen PSI, Switzerland} 

\author{B. Roessli}
\affiliation{Laboratory for Neutron Scattering ETHZ \& Paul Scherrer
Institut, CH-5232 Villigen PSI, Switzerland} 

\author{S. Kojima}
\affiliation{Institute of Materials Science, University of Tsukuba, Tsukuba, Ibaraki 305-8573, Japan}

\author{S.G. Lushnikov}
\affiliation{Ioffe Physical Technical Institute, 26 Politekhnicheskaya,
194021, St. Petersburg, Russia}

\author{P. G\"unter}
\affiliation{Institute for Quantum Electronics, ETH H\"onggerberg-HPF, CH-8093 Zurich, Switzerland}

\date{\today} 

\begin{abstract} 

\pagenumbering{arabic}

\noindent We show that the neutron diffuse scattering in 
relaxor ferroelectric \pznpt (x=0.07) consists of 
two components. The first component is strictly elastic but extended in 
q-space and grows below 600~K. The second component, that was not reported before for the \pznpt relaxor 
ferroelectrics, is quasi-elastic with a line-width that has a similar temperature dependence as the width of the 
central peak observed by Brillouin spectroscopy. The temperature dependence of the 
susceptibility of the quasi-elastic scattering has a maximum at the ferroelectric transition.
\end{abstract}
\maketitle
\noindent
Complex perovskites with the chemical formula PbB$_{1/3}$Nb$_{2/3}$O$_3$ (B = Mg, Zn) are important 
materials for applications because they possess giant piezoelectric constants when doped 
with PbTiO$_3$ (PT). The temperature-concentration phase diagrams of \pmn (PMN) and \pzn (PZN) 
are complex. The average crystal structure of pure PMN is cubic at all temperatures. 
For small concentrations of PT there is a phase transition from cubic to rhombohedral 
symmetry at low temperature. 
Upon increasing the amount of PT a morphotropic phase boundary  that separates 
the rhombohedral phase from a phase with tetragonal 
symmetry appears at x=0.32~\cite{shuvaeva} for \pmnpt and x=0.08~\cite{lima} for \pznpt. 
The dielectric permittivity of PMN and PZN has a broad maximum around 265~K and 320~K 
respectively.  At higher temperatures the dependence of the refractive index of both 
materials deviates from the expected linear dependence and this was explained by the appearance 
in the crystals of polar regions of nanometer size (PNR)~\cite{burns}. 
Evidence for the formation of PNR in \pmnpt and \pznpt comes from the presence of 
temperature-dependent diffuse scattering (DS) close to the Bragg reflections observed by both X-rays and neutron diffraction. 
Upon approaching T$_c$, the intensity of the diffuse scattering increases which 
suggests that the PNR's grow and that the ferroelectric state in the relaxor ferroelectrics is eventually reached 
when the PNR's produce a spontaneous polarization in the same direction. Therefore it is of importance to understand  
the structure and dynamics of the PNR's. The local structure of the PNR's has still not be completely 
determined and whether or not the local symmetry of the PNR's 
is lower then cubic already at high temperature or undergoes a local phase transition 
above or at T$_c$ is a matter of debate~\cite{egami_prl_2005,roth_prl_2007}.  
The diffuse scattering measured by neutron scattering in \pmnpt is not entirely static but has an additional  
quasi-elastic component (QE) that corresponds to the dynamics of the PNR's~\cite{gva_prb,hiraka}. For PMN 
the intensity associated with the QE scattering was found to follow the temperature dependence 
of the dielectric permittivity~\cite{gva_jpcm_pmn}. The central peak (CP) and 
the QE component are also observed in doped PMN by neutron scattering~\cite{gva_jpcm_pmnpt}, although the CP 
intensity is weaker then in pure PMN. In \pmnpt with x=0.32 the susceptibility of the QE scattering increases 
below the Burns temperature and has a broad maximum at the temperature where the cubic-to-tetragonal phase transition 
occurs~\cite{gva_jpcm_pmnpt}.  

\begin{figure}[h]
  \includegraphics[width=0.4\textwidth]{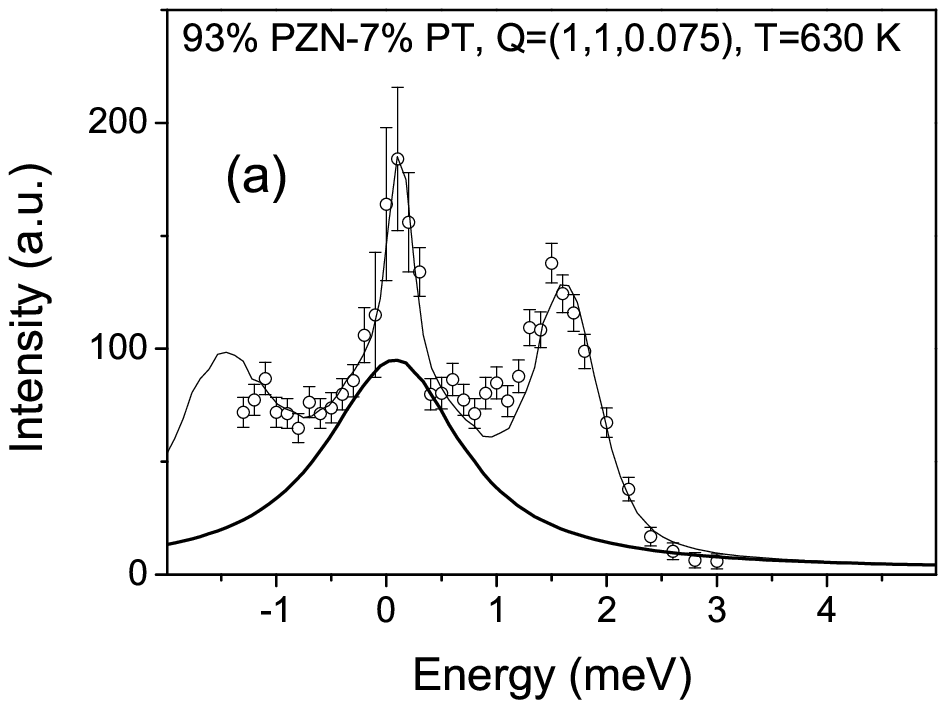}
  \includegraphics[width=0.4\textwidth]{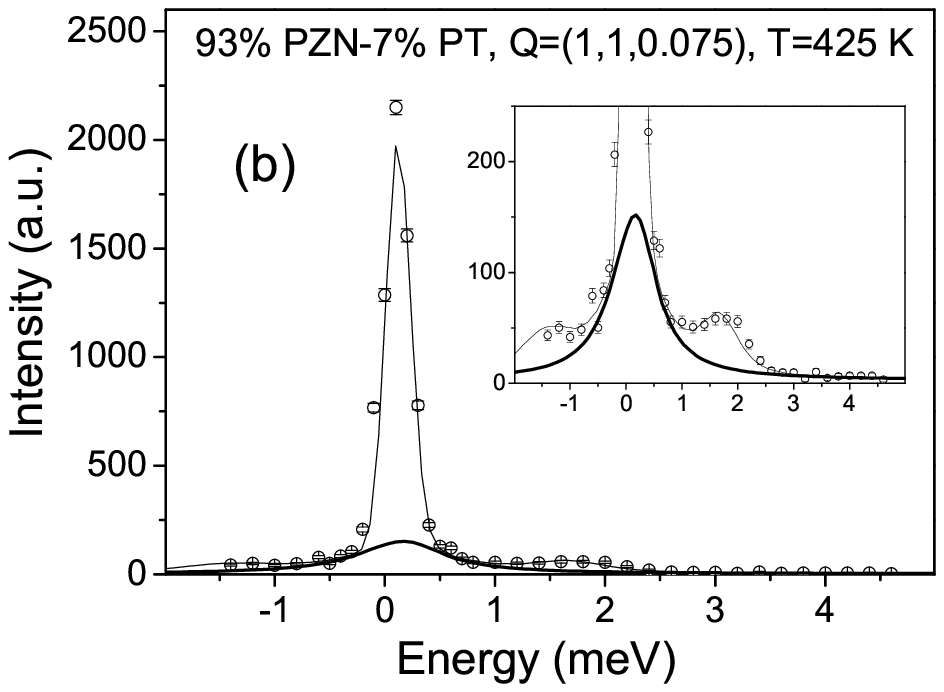}
  \caption{Typical const-Q scans at T~=~630~K ({\bf a}) and at T~=~425~K ({\bf b}). The solid line is the fit. 
           The bold line emphasizes the contribution of the QE component. 
           }
\label{fig1}
\end{figure} 

\begin{figure}[h]
  \includegraphics[width=0.49\textwidth]{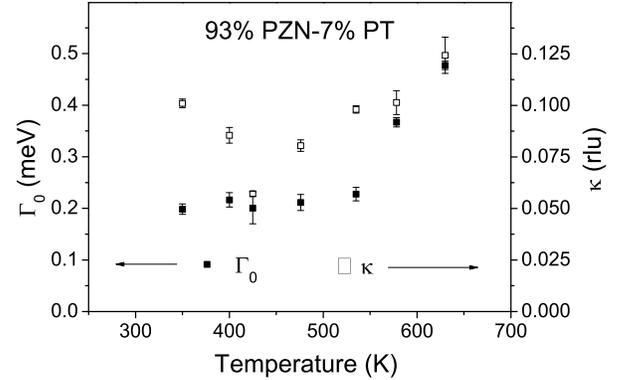}
  \caption {The temperature dependences of the damping of the 
           QE scattering $\Gamma_0$ and of the inverse correlation length $\kappa$. 
           }
\label{fig2}
\end{figure}

\pmnpt and \pznpt solid solutions have a comparable $(x,T)$-phase diagrams 
and the dielectric response of both undoped materials show relaxor properties. In PMN and in PZN, it was shown 
by neutron scattering that the shape and the temperature dependence of the DS are similar, and 
it was proposed in Refs.~\cite{gva_jpcm_pmn,stock_prb,gva_jpcm_pmnpt} that a random-field model with cubic anisotropy 
could describe the physical properties of both families of compounds. 
In the attempt to propose a unified description of \pmnpt and \pznpt relaxors, an unsolved question 
is the behavior of the QE part of the diffuse scattering. 
Hlinka et al. found that the DS in \pznpt with x=0.08 is truly elastic, with any dynamics being slower 
then 8 GHz~\cite{hlinka_jpcm}. This result contrasts with those obtained by 
neutron scattering from PMN and \pmnpt (x=0.32) where the quasi-elastic component has a lifetime varying between 
0.05-0.2 THz as a function of the temperature~\cite{gva_prb,hiraka,gva_jpcm_pmn,gva_jpcm_pmnpt}, 
as well as results from Brillouin experiments in \pznpt\cite{kojima}.

In this Letter we present neutron scattering results obtained in \pznpt with x=0.07. We show that the low-energy part of the 
excitation spectrum  contains in addition to the elastic DS observed by Hlinka et al. 
a QE component that was not reported before. We present an analysis of the evolution of the intensity and of 
the line width of this component as a function of the temperature and compare the results with those obtained 
in \pmnpt (x=0 and x=0.32)~\cite{gva_jpcm_pmn,gva_jpcm_pmnpt}.  

\begin{figure}[h]
  \includegraphics[width=0.49\textwidth]{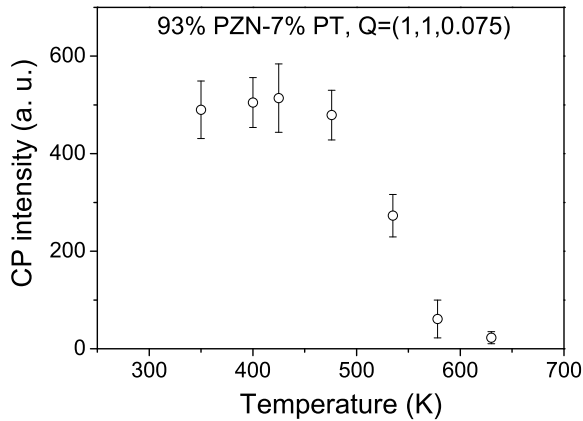}
  \caption{The temperature dependences of the elastic component of the diffuse scattering. 
           }
\label{fig3}
\end{figure}

\begin{figure}[h]
  \includegraphics[width=0.49\textwidth]{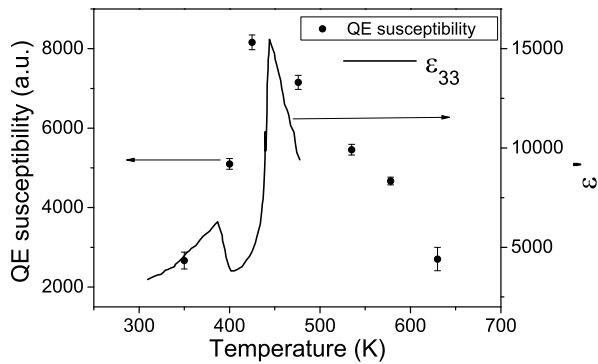}
  \caption{The temperature dependences of the susceptibility $\chi(0,T)$ of the QE scattering and of the dielectric 
           constant $\varepsilon'$ taken from Ref.~\cite{hosono}. 
           }
\label{fig4}
\end{figure}

\noindent We used the cold neutron 3-axis spectrometer TASP~\cite{tasp} (SINQ~\cite{sinq}, PSI) operated in the constant 
k$_f$-mode with k$_f = 1.64$~\AA $^{-1}$. A PG filter was installed in the scattered beam to reduce 
contaminations by higher-order wavelengths. The horizontal collimation was $10'$/\AA $-80'-80'-80'$.
With that setting the energy resolution was $0.26$~meV. The sample ($\sim1$~cm$^3$ in volume) was 
mounted in a standard furnace with the $<1~\bar{1}~0>$ 
cubic axis vertical. Both constant-Q and constant-energy scans were performed 
around the (1~1~0) Bragg reflection in the temperature range 300~K $<$ T $<$ 630~K. Fig.~\ref{fig1}a shows a typical 
energy scan performed at T= 630~K and $\vec Q=(1,1,0.075)$. The spectrum consists of a transverse acoustic (TA) phonon and 
of scattering centered around the elastic position. The incoherent scattering was measured at high temperature 
at $\vec Q=(1,1,0.3)$ and subtracted from the data. Fig.~\ref{fig1}b shows the same scan at T=425~K, i.e.  
close to the phase transition to the ferroelectric state. At this temperature the neutron intensity 
measured around the elastic position has strongly increased. The lineshape of the TA phonon is 
well described by a damped-harmonic-oscillator function convoluted with the resolution function of the spectrometer. 
We obtain for the TA dispersion $\omega_{TA}=d\sin(\pi q)$ with $d=6.8\pm0.1$~meV. 
It turns out from the analysis of the data at all temperatures the 
central component cannot be reproduced by a resolution-limited Gaussian function only and consists of two components,
the truly elastic diffuse scattering observed by Hlinka et al.~\cite{hlinka_jpcm} and quasi-elastic scattering. 
The QE scattering is modeled by a Lorentzian function as 
was done before in PMN~\cite{gva_prb,hiraka,gva_jpcm_pmn} and \pmnpt~\cite{gva_jpcm_pmnpt}:
\begin{equation}
	\chi''_{QE}(\omega,q,T)={{\chi(0,T)}\over{1+(q/\kappa)^2}}{{\omega\Gamma_q}\over{\Gamma^2_q+\omega^2}}, \nonumber
\end{equation}
where $\Gamma_q=\Gamma_0+Dq^2$. We find that $D=18\pm1$~meV\AA $^{2}$ in \pznpt (x=0.07). 
$\kappa=1/\xi$ is the inverse of the correlation length $\xi$ and is obtained from constant energy-scans. 
Fig.~\ref{fig2} shows the temperature dependence of $\kappa (T)$ and $\Gamma_0(T)$;  
$\kappa\sim$ 0.12 $\pm0.01$~\AA $^{-1}$ at 630~K and  
decreases with decreasing temperature as does the line width of the QE scattering: $\Gamma_0=0.48\pm0.01$ meV at T=630~K and 
$0.21\pm0.01$ meV below $\sim475$~K, where the dynamics of the PNR's freeze. A similar behavior 
is observed for the temperature dependence of the width of the central peak in light scattering where $\Gamma_0\sim$80 GHz 
($\sim0.3$~meV) in the ferroelectric phase~\cite{kojima}. On the other hand, the intensity of the elastic part of 
the diffuse scattering increases strongly below 600~K and reaches a maximum around T$_c$ as shown in Fig.~\ref{fig3}. 
The temperature dependence of the susceptibility of the QE scattering is shown in Fig.~\ref{fig4}. $\chi(0,T)$ increases 
strongly on cooling below 630~K, reaches a maximum around $T_C$ 
and decreases rapidly in the ferroelectric phase. The temperature dependence of the susceptibility of the 
QE scattering in \pznpt (x=0.07) as a similiar temperature dependence as the dielectric permittivity like in \pmnpt (x=0,0.32). 


To conclude,  we found that in \pznpt (x=0.07) the DS consists of two components associated with two different timescales, which  
reconciles the discrepancy between light scattering~\cite{kojima} and 
previous neutron scattering measurements~\cite{hlinka_jpcm}. We have also shown that the temperature dependences of the CP 
and QE components have a similar behavior in both \pmnpt and \pznpt (x=0.07), which brings further evidence that random fields 
play an important role in the physics of the relaxor ferroelectrics, as discussed previously~\cite{gva_jpcm_pmn,stock_prb,gva_jpcm_pmnpt}. 

\begin{acknowledgments}
The authors thank R.A. Cowley for discussions. This work was performed at the spallation neutron 
source SINQ, Paul Scherrer Institut, Villigen (Switzerland) and was partially supported by the Swiss National Foundation 
(Project No. 20002-111545). 
\end{acknowledgments}

\end{document}